\documentclass[twocolumn,english,aps,prl,showpacs,groupedaddress]{revtex4}
\usepackage{subfigure}
\usepackage{graphics}

\usepackage{graphicx}
\usepackage[T1]{fontenc}
\usepackage[latin9]{inputenc}
\usepackage{ifsym}
\usepackage{wasysym}
\usepackage{babel}

\begin{document}

\title{Characteristic  spatial scale of vesicle pair interactions in a plane linear flow}

\author{Michael Levant, Julien Deschamps$^1$, Eldad Afik, and Victor Steinberg}

\affiliation{Department of Physics of Complex Systems, Weizmann
Institute of Science, Rehovot, 76100 Israel
$^1$ Permanent address:IRPHE, Aix-Marseille Univ., 13384 Marseille, France; IRPHE, UMR
CNRS 6594, 13384 Marseille, France }

\date{\today}

\begin{abstract}
We report the experimental studies on interaction of two vesicles trapped in a micro-fluidic four-roll mill, where a plane linear flow is realized. We found that the dynamics of a vesicle in tank-treading motion is significantly altered by the presence of another vesicle at separation distances up to $3.2\div 3.7$ times of the vesicle effective radius. This result is supported by measurement of a single vesicle back-reaction on the velocity field. Thus, the experiment provides the upper bound for the volume fraction $\phi=0.08\div 0.13$ of non-interacting vesicle suspensions.
\end{abstract}

\pacs{87.16.D-, 82.70.Uv, 83.50.-v}

\maketitle

  The coupling between rheological, macroscopic properties of
complex fluids, such as suspensions, emulsions, polymer solutions etc, and the microscopic dynamics of the deformable micro-objects immersed in them has been a long-standing problem in physics, chemistry, and engineering. The simpler the micro-object dynamics, the more impressive the progress in understanding rheology of the corresponding complex fluid. A prominent example is a colloid suspension, where the correction for the effective viscosity of a dilute suspension due to immersed solid spheres has been obtained about a century ago \cite{landau}. However, for such complex fluids, as suspension of vesicles, capsules or red blood cells, which obviously have great biological and industrial applications, the relation between their rheological properties and the dynamics of the deformable micro-objects still remain under discussion due to  their more elaborate microscopic behavior.
 For vesicle suspensions, the currently available theories \cite{misbah,vlahovska,misbah2,vergeles} deal only with the dilute regime, so that its rheology is described in the limit of a single vesicle by completely neglecting vesicle interactions. On the other hand, the results of the two currently available experiments on the dependence of the effective viscosity of the vesicle suspension $\bar{\eta}$ on the vesicle viscosity contrast between the inner and outer fluids $\lambda$ contradict each other in measurements taken at about the same volume fraction of vesicles $\phi$ of 0.11-0.12 \cite{kantsler,vitkova}. Though one of the experiments \cite{vitkova} agrees with the theory \cite{vlahovska,misbah2}, the other \cite{kantsler} shows a non-monotonic dependence of $\bar{\eta}$ on $\lambda$, contrary to the predictions for dilute suspension \cite{misbah,vlahovska,misbah2,vergeles}. The results of the experiments reported in Ref. \cite{kantsler} indicate that hydrodynamically-assisted vesicle interactions could be a cause for the observed qualitative difference with the predicted behavior. Indeed, both vertical velocity fluctuations and vesicle inclination angle in tank-treading (TT) motion in a shear flow were significantly affected by many surrounding vesicles, as measurements presented in Ref. \cite{kantsler} showed. In this respect, the key question is whether a long-range hydrodynamically-assisted interaction of vesicles in flow can qualitatively modify the rheological properties of the suspension and at what characteristic length scale does the vesicle interaction become significant to achieve this.  To the best of our knowledge, the role of long-range hydrodynamic interaction was addressed theoretically \cite{batchelor} and numerically \cite{peyla} only for colloids, where interaction of two suspended spheres in a shear flow was studied. \\
 \begin{figure}
\includegraphics[width=0.9\columnwidth]{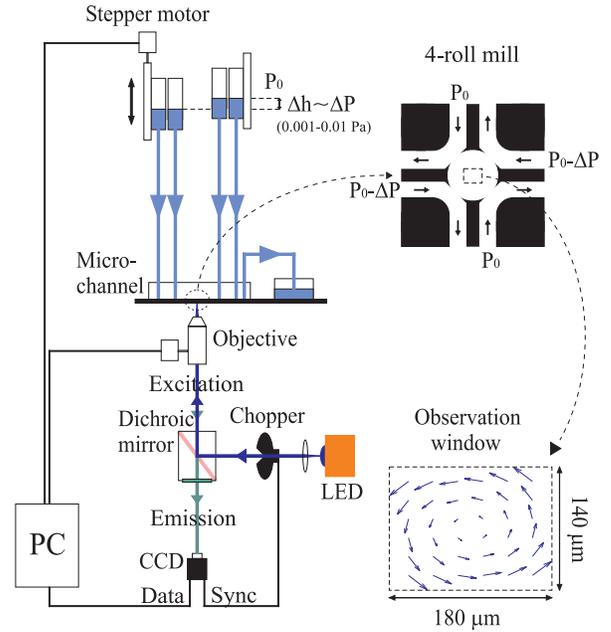}

\caption{(Color on line) Experimental setup.}

\label{fig:setup}
\end{figure}
\indent Here we report the studies of the correlation and the range of the interaction of two vesicles in a plane linear flow. By applying a novel and unique technique for vesicles, deviations of the vesicle inclination angle in the TT motion from the stationary value were used as a sensitive quantitative detector of the interaction strength between vesicles. It provides the characteristic length scale of the long-range hydrodynamically-assisted vesicle pair interaction, which determines a lower bound of the interaction scale for a semi-dilute vesicle suspension. This method could in principle be applied to other deformable objects.\\
\begin{figure}
\includegraphics[width=0.9\columnwidth]{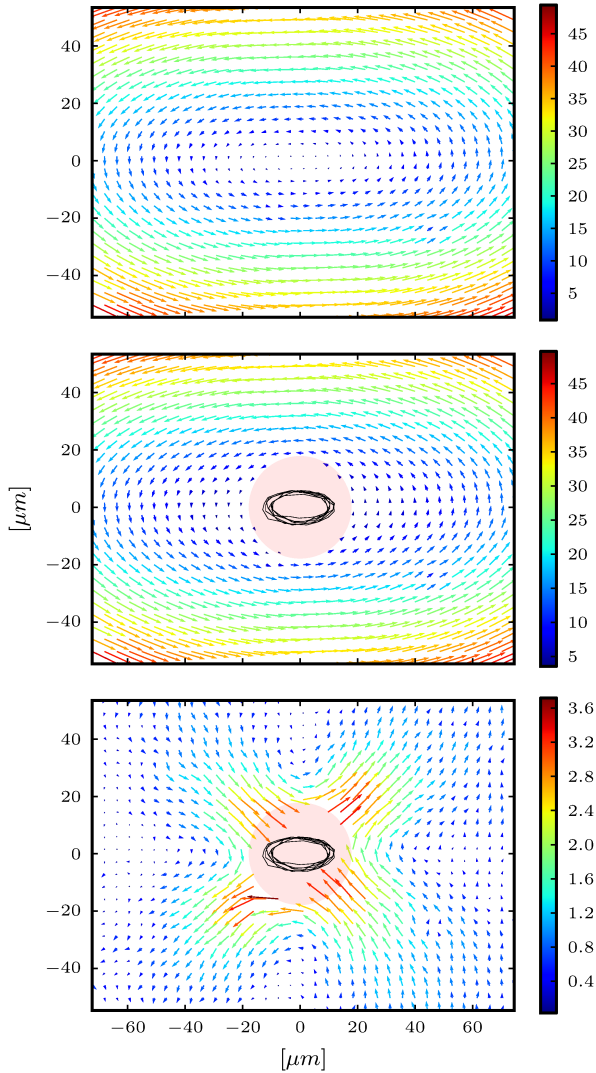}

\caption{ (Color) The effect of a nearly spherical vesicle on the velocity field. The 2D
time-averaged velocity fields (over $122 s$ on 7 vesicle orbits) are presented: : (top) reference field without a vesicle; (middle) in the presence of a nearly
spherical vesicle ($R \simeq 17 \mu m$); (bottom) difference field, i.e. subtraction of the reference field from that measured in the presence of the vesicle.
Colors indicate velocity magnitude $[\mu m\cdot s^{-1}]$;
 Vesicle trajectory is denoted in black.
The plotted circle shows the vesicle size and mean position. Note that velocity differences exceeding 1 $\mu$m/s are significant. It was verified by comparing two reference flow fields.}

 \label{fig:Flow field difference}
\end{figure}

\begin{figure}
\includegraphics[width=0.9\columnwidth]{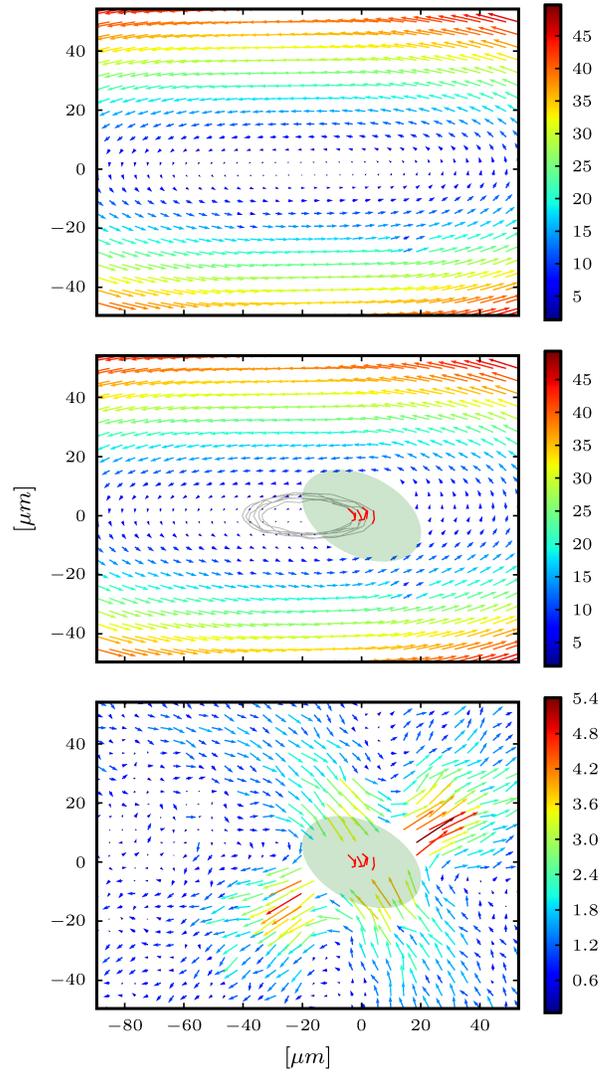}

\caption{(Color) The effect of a deflated (non-spherical) vesicle on the velocity field. The 2D partial
time-averaged velocity fields are presented; in the presence of the vesicle the averaging is taken over $2 s$ segments (marked in red) along each of 6
vesicle orbits (in black): (top) reference field without a vesicle; (middle) in the presence of a deflated vesicle ($R \simeq 16.6 \mu m$, $\Delta\simeq 0.39$); (bottom) difference field, i.e. subtraction of the reference field from that measured in the presence of the vesicle. Colors indicate velocity magnitude $[\mu m\cdot s^{-1}]$. The plotted ellipse shows the vesicle size and mean position during the time over which the averaging was taken. Note that velocity differences exceeding 1 $\mu$m/s are significant. This was verified by comparing two reference flow fields.}

 \label{fig:Flow field difference3}
\end{figure}

\indent A vesicle is a drop of fluid, encapsulated in a lipid bilayer membrane and
suspended in either the same or different fluid. The impermeability and the
inextensibility of the lipid membrane dictate conservation of both
volume and surface area of the vesicle. Vesicles undergo a TT motion in a planar
linear flow at $\lambda=1$ and $\omega/s<2$ with $0<\Delta<1$ \cite{deschamps}.  Here $\Delta=A/R^2-4\pi$ is the excess area, $R=(3V/4\pi)^{1/3}$ is  the vesicle effective
radius, $V$ is the vesicle volume, $A$ is the vesicle surface area, and $\omega$ and $s$ are the vorticity and strain rate, respectively \cite{lebedev}. At a fixed viscosity ratio, the
inclination angle $\theta$ decreases with increasing $\Delta$ and increasing
$\omega/s$ \cite{lebedev,kantsler1,kantsler2,norman}.\\
\indent The experiments were conducted in a micro-fluidic four-roll mill \cite{muller,deschamps}, implemented in silicone
elastomer (Sylgard 184, Dow Corning) by soft
lithography. The micro-channel was kept in a vacuum chamber for $\geq 1$
hour before filling with vesicle solution, to avoid air bubbles. The key component of this device is a dynamical trap,
which allows a long observation time compared with the orbit period of vesicles in the flow. The observation time is limited by the fluorescent lipids bleaching. The flow was driven by gravity and the control parameter $\omega/s$ was varied continuously by variation of the pressure drop $\Delta P$ across the device in the range [0-0.01] Pa (Fig. \ref{fig:setup}).

\begin{figure*}
\centering
\includegraphics[width=\textwidth]{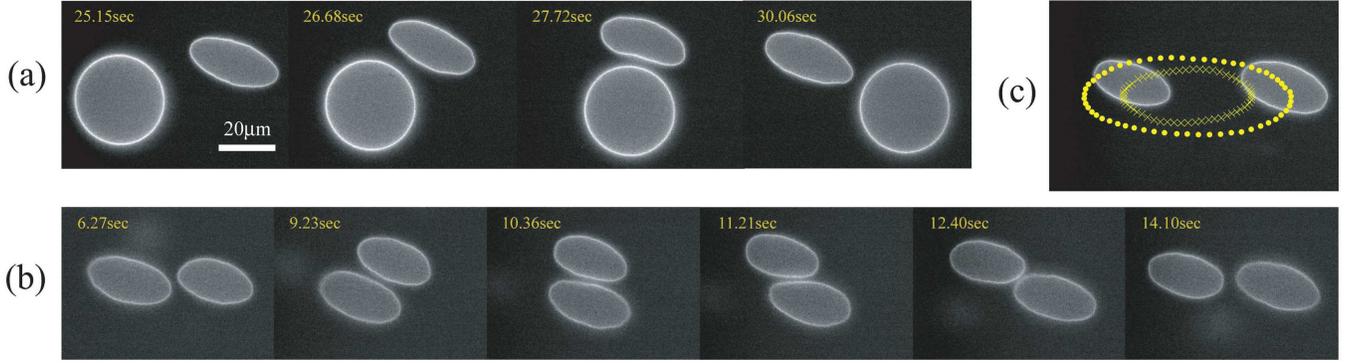}
\caption{(Color on line) Snapshots of vesicle pair interactions. a)
$\Delta\simeq0,1,\;R\simeq13.2,9.2 \mu m,\;\omega/s=1.33$. b)
$\Delta\simeq0.88,0.77,\;R\simeq10.1,11.9 \mu m,\;\omega/s=1.52$. c)
Trajectories of interacting vesicles inside the dynamical trap at
$\omega/s=1.52$} \label{fig:snaps}
\end{figure*}

 Vesicles were prepared in water
($\eta_{out}=\eta_{in}=1mPa\cdot sec$) via electro-formation
\cite{electroformation}, using lipid solution, consisting of 85\%
DOPC lipids (Sigma) and 15\% NBD-PC fluorescent lipids (Molecular
Probes), dissolved in 9:1 v/v chloroform/methanol \cite{kantsler1,kantsler2}. The electro-formation cell consisted of two Indium Tin Oxide (ITO)
coated glass electrodes with a 1 mm teflon spacing between them.
Droplets of the lipid solution were spread on the glasses. The
solvent quickly evaporated leaving just the dry lipids. The cells
were then filled with de-ionized water (Merck, R> 1M =cm). An
external AC voltage, sine wave, 3 volts peak-to-peak, 10 Hz, was
applied to the electrodes. The electro-formation cells were placed
on an orbital shaker, rotating at 90 rpm for $\geq 8$ hours. New
solution was prepared prior to each experiment \cite{kantsler1,kantsler2}. The vesicles
dynamics were monitored using inverted fluorescence microscope
(IMT-2, Olympus) inside a $180\times140\mu m$ observation window,
parallel to the flow (Fig. \ref{fig:setup}). The images were collected using a Prosilica EC1380
CCD camera, aligned with the shear axis, at 30 fps, with a spatial
resolution of 0.27 $\mu m/pixel$. The camera was synchronized with a
mechanical chopper on the path of the excitation beam to reduce
exposure time. Vesicles were loaded into the dynamical trap  via an inlet channel at the mid-plane of
the micro-channel (of 300 $\mu$m height). To isolate just two vesicles near the trap center, we repeatedly varied $\Delta P$ quickly (on the order of a second) generating short elongation flow pulses (open trajectories), which expelled the vesicles located far from the trap center. After these perturbations and setting the fixed value of $\omega/s$, we allow the flow to relax during several vesicle relaxation times, which is of the order of 5-10 s \cite{norman}, before starting the measurements.  Moving along closed trajectories imposed by
the external flow field, the vesicles were repeatedly approaching
and moving away from each other (Fig. \ref{fig:snaps} c and movie in \cite{sm}). A special
attention was given to having the largest cross-sections of the
vesicles in the same plane in order to reduce 3D
interaction effects. Image processing was based on a binary recognition of each vesicle used to
extract the coordinates of its center. The relative position of the
membrane was determined in the frame of reference of the vesicle
using intensity variations along the radial directions. The
subsequent analysis included elliptical approximation of the shape
and determination of the lengths of the main axes, the inclination
angle, and the center of the ellipse. Following \cite{kantsler1,kantsler2}, both $R$ and $\Delta$ were calculated from the main axes. \\
\indent Before presenting our main results regarding the interaction of two vesicles, it is insightful to examine direct measurements of the velocity field in the vicinity of a single vesicle, demonstrating the back-reaction of the vesicle on the flow. The velocity measurements were conducted by micro Particle Image Velocimetry ($\textrm{\ensuremath{\mu}}$PIV). In this experiment an aqueous vesicles suspension was mixed with fluorescent particles
(Fluoresbrite\textsuperscript{\textregistered} YG Microspheres $0.5\mu m$
from Polysciences, Inc.) at $2.6\times10^{7}particles\cdot\mu l^{-1}$,
allowing imaging of the flow in the four-roll mill trap. Images were
recorded at 90.9 fr/s rate on a computer using the Motmot
Python camera interface package \cite{Straw_Dickinson_2009}, controlling a GX1920 camera from Allied Vision Technologies. The
images were then filtered (Laplace filter using Gaussian second derivatives
\cite{scipy}) and processed using Gpiv \cite{gpiv} at interrogation windows of $32\times32$ pixels (corresponding to about $8.6\times8.6\mu m^{2}$) with $50\%$ overlap.
The effect of a vesicle with vanishingly small $\Delta$,
located near the stagnation point of the trap and moving on a small orbit in a plane linear flow with $\omega=0.46$ s$^{-1}$ and $\omega/s=1.8$, on the velocity field was measured. Figure \ref {fig:Flow field difference} at the bottom presents the difference velocity field, which results from a subtraction of the reference field (Fig.\ref{fig:Flow field difference} at the top) from that measured in the presence of the vesicle (Fig. \ref{fig:Flow field difference} at the middle). It allows to study the back-reaction of the vesicle on the flow, and to estimate
the effective distance of the back-reaction. The most pronounced feature of the field difference is the alternation of
the strain strength in the vicinity of the vesicle, which is significant at
scale of up to about $3R$ along the strain main axes. The compression and
stretching directions of the field difference show sign inversion compared to those
found in the reference flow field, where a vesicle is absent (Fig. \ref{fig:Flow field difference} at the top). This can be attributed to the vesicle's volume and area conservation,
opposing extension and compression. It can be also viewed as a result
of the tank-treading motion of the membrane, which must move at constant
tangential velocity. Indeed, it was also observed in a snapshot of velocity field around a vesicle in 2D numerical simulations \cite{seifert}. Very similar image of the disturbance velocity field around a solid sphere in a shear flow was recently obtained by numerical simulations \cite{peyla}.\\
\indent Similar analysis of the difference velocity field around a deflated (non-spherical) vesicle moving on a small orbit in a plane linear flow with $\omega=0.46$ s$^{-1}$ and $\omega/s=1.4$  after partial time averaging on small part of its  six orbits as shown in the vesicle center is presented in Fig. \ref{fig:Flow field difference3}. The pronounce difference with Fig. \ref{fig:Flow field difference} is symmetry breaking of compression and stretching directions in the disturbance velocity field: the stretching direction deviates from the strain direction by about $\pi/12$.\\
\begin{figure}
\includegraphics[width=7.5cm]{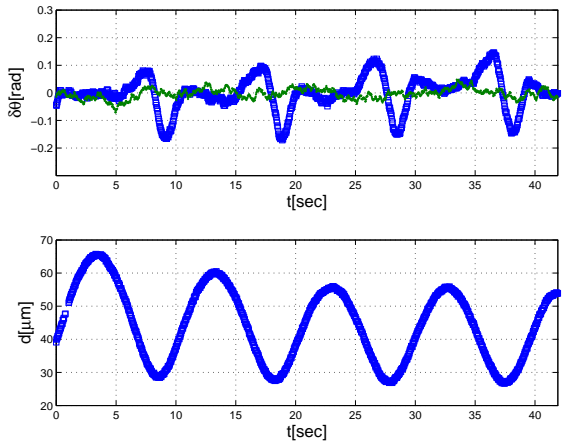}
\caption{(Color on line) The upper plot shows a time series of $\delta\theta$, the deviations of the vesicle inclination angle from its stationary value (blue squares), for a deflated vesicle along its orbit as it interacts with a spherical one. The green dots denote $\delta\theta$ of the same deflated vesicle after removal of the spherical vesicle. The lower plot shows a time series of the distance between the centers of the interacting vesicles.}
\label{fig:timeseries}
\end{figure}
\indent We conducted two types of observations of vesicle pair
interactions: interaction of spherical and deflated vesicles in a pair and  interaction of two deflated vesicles in a pair.
First we observed interaction of a deflated vesicle
in the range of parameters $\Delta\in[0.7,1]$ and $R\in[6,11]\mu m$,
placed near the stagnation point of the flow, and a vesicle with
vanishingly small $\Delta$ and $R\in[11,17]\mu m$ rotating around the former, all together nine
such pairs. By choosing one spherical vesicle in each pair, we reduce the number of parameters and simplify quantification of the problem (see Fig. \ref {fig:snaps} a and movie in Ref. \cite{sm}). We took sufficiently long time series to observe several periods of the vesicle interaction. The latter is characterized by $\delta\theta=\theta-\langle\theta\rangle$, the deviation of the inclination angle of the
deflated vesicle $\theta$ from its mean stationary value $\langle\theta\rangle$ \cite{kantsler1,kantsler2} and is typically accompanied by
the vesicle deformation. The observed $\delta\theta$ due to the vesicle interactions are up to one order of magnitude larger than due to thermal fluctuations.\\
\begin{figure}
\includegraphics[width=8cm]{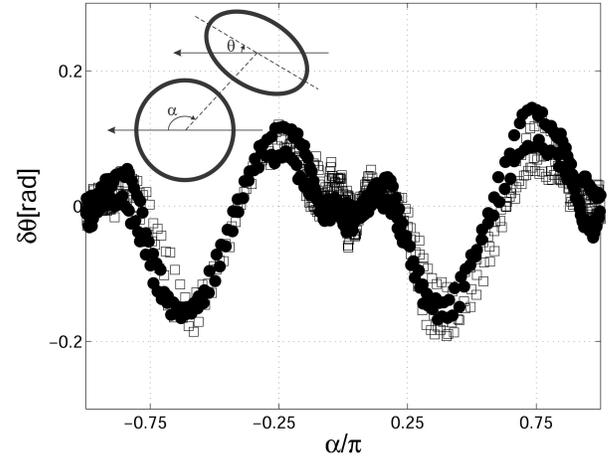}
\caption{$\delta\theta$ versus $\alpha/\pi$. Two
data-sets are for
$\Delta\simeq0,1,\;R\simeq13.2,9.2\;\mu m,\;\omega/s=1.33$ - circles, and
$\Delta\simeq0,0.82,\;R\simeq12.5,7.5\;\mu m,\;\omega/s=1.27$ - squares. }
\label{fig:angles}
\end{figure}
\indent To relate the deviations of the vesicle inclination angle from its stationary value $\delta\theta$ for a deflated vesicle along its orbit, as it interacts with a spherical one, to the distance between vesicle centers we added two plots in Fig. \ref{fig:timeseries}. To determine $\delta\theta$ as a function of time for the same vesicle, $\theta(t)$ of the same deflated vesicle was measured in the presence of a spherical one and then after its removal (Fig. \ref{fig:timeseries} at the top). At the same time, the distance between the centers of the interacting vesicles was measured (Fig. \ref{fig:timeseries} at the bottom).\\
\indent In order to relate the data to the structure of the strain rate of the velocity field around a vesicle (Fig. \ref{fig:Flow field difference}), we plot $\delta\theta$ versus $\alpha$, the angle between the line connecting both vesicle centers and the horizontal direction (see inset in Fig. \ref{fig:angles}). In the two different data sets presented in
Fig. \ref{fig:angles}, where each data set contains 3-4 interaction
events, we observe a repetitive pattern, periodic in $\alpha/\pi$.
The pattern is characterized by maxima of $\delta\theta$
occurring in the stretching direction of the velocity field difference at
$\alpha/\pi=-1/4$ and $3/4$, while $\delta\theta$ minima
are shifted from the compression direction at $\alpha/\pi=-3/4$ and
$1/4$ (see Fig. \ref{fig:Flow field difference}). The deep minima of $\delta\theta$ result from a short-range vesicle-vesicle interaction, which is accompanied by strong vesicle deformation (see Fig. \ref{fig:snaps} a) at distance $d/R\approx 1$ and bounded by a width of a fluid lubricating layer, where $d/R$ is the normalized distance between vesicle centers, and $R$ is
the radius of the nearly spherical vesicle.
 The short-range interaction was studied experimentally for a vesicle \cite{kantsler} and droplet \cite{guido} pairs and numerically for a capsule pair \cite{lac}. The maxima of $\delta\theta$ occur as the result of a long-range hydrodynamically-assisted vesicle-vesicle interaction at $d/R$ determined above. It has not been studied before for any soft objects and appears to be the main interest of our investigation here.\\
\begin{figure}
\includegraphics[width=7.5cm]{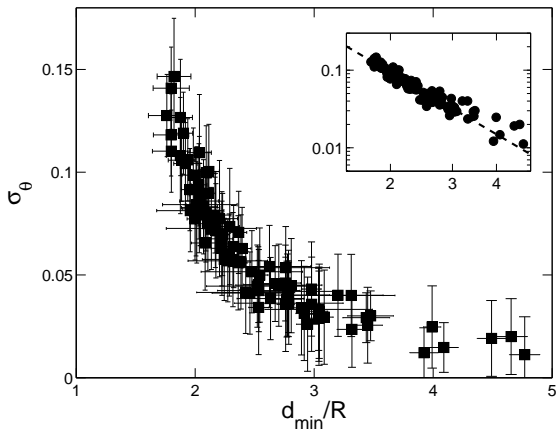}
\caption{\small{$\sigma_{\theta}$ versus $d_{min}/R$. The data are taken over many cycles in time series of 9 vesicle pairs. Inset: The same data on logarithmic scale. Dash line is the fit
$\sigma_{\theta}\sim (d_{min}/R)^{\beta}$ with $\beta=-2.65\pm0.2$. (see also Fig. 1 in \cite{sm} where different vesicle pairs are shown by different symbols)}}
\label{fig:D0}
\end{figure}
\indent Figure \ref{fig:D0} presents $\sigma_{\theta}$,
the rms of the deflated vesicle inclination angle $\theta$ (inset in Fig. 5) as a function of  $d/R$ in the vicinity of its minimal value $d_{min}/R$, the relevant parameter in the problem. As seen in the plot, $\sigma_{\theta}$ reduces with $d_{min}/R$ and reaches background noise level of the order $0.02\div0.03$ rad at $d_{min}/R\approx3.2\div3.7$ (in Fig. 1 of \cite{sm} the same data are shown, and different vesicle pairs are presented by different symbols). This noise level, measured independently for a single vesicle in the
four-roll mill, is attributed to thermal fluctuations and
non-uniformities of the flow (see Fig. \ref{fig:timeseries}). The same data, shown in the inset in Fig. \ref{fig:D0} on a logarithmic scale, is fitted by $\sigma_{\theta}\sim(d_{min}/R)^\beta$ with $\beta=-2.65\pm0.2$. It should be compared with a long range hydrodynamic particle-particle interaction in a shear flow, where the presence of a second particle changes the induced strain rate of a test particle as $d^{-3}$ for unbounded and $d^{-2}$ for bounded suspensions \cite{batchelor,peyla}.
The spatial scales, at which the vesicle pair interaction reduced down to the noise level, corresponds to a vesicle volume fraction of $\phi\approx0.08\div0.13$. \\

\begin{figure}
\includegraphics[width=\columnwidth]{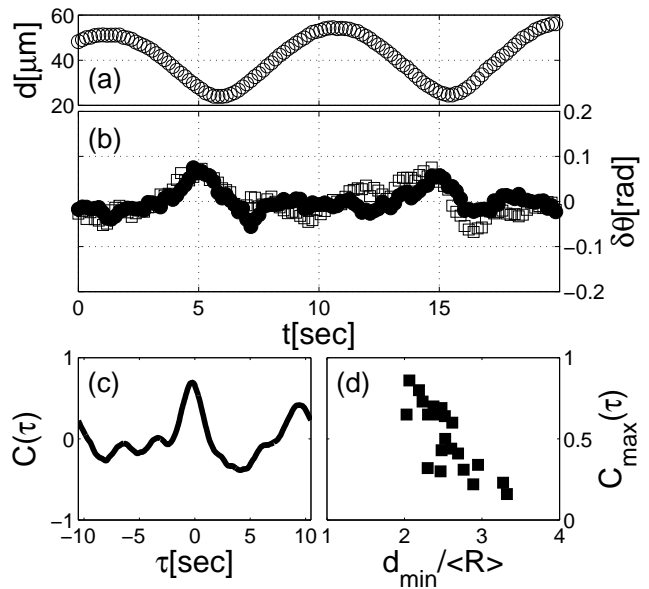}
\caption{\small{Characteristic dynamics of two deflated vesicle pair,
$\Delta=0.876,0.737,\;R=10.3,10.18\mu m$. a) $d$ versus time. b) $\delta\theta$ versus time.  c) $C(\tau)$ versus $\tau$. d) $C_{max}(\tau)$ versus $d_{min}/\langle R\rangle$.}} \label{fig:dynamics}
\end{figure}
 \indent In a second, general case of two interacting deflated vesicles (see Fig. \ref{fig:snaps} b,c) more parameters characterizing both the vesicle and flow geometry (see Fig. \ref{fig:Flow field difference3}) are involved, which makes the situation less quantitatively tractable.
In this case a correlation of the inclination angles of
the vesicles near a local minima of the distance between vesicle centers was observed (Fig. \ref{fig:dynamics} a,b). We investigated eight pairs of vesicles with similar
physical parameters in the range $\Delta\in[0.5,0.9],\;R\in[6,12]\mu
m$ and found high correlation, measured via the peak $C_{max}(\tau)$ of the
normalized cross-correlation function $C(\tau)=\langle{\delta\theta_1(t)\delta\theta_2(t+\tau)}\rangle_{t}/{\sigma_{\theta_1}\sigma_{\theta_2}}$ for nearly colliding
vesicles (see Fig. \ref{fig:dynamics} c), followed by its rapid decrease in the region of $d_{min}/\langle
R\rangle\approx2\div3.3$ (Fig. \ref{fig:dynamics} d). Here $\langle R\rangle$ is the arithmetic average of the radii of the two vesicles in pair. Thus the characteristic length scale defined from Fig. \ref{fig:dynamics} d, where $C_{max}(\tau)$ reaches the noise level, is $d_{min}/\langle
R\rangle\approx 3.3$ that corresponds to a volume fraction of
$\phi\approx 0.12\pm0.03$.\\
\indent To conclude, we have found that the
dynamics of a single vesicle can be significantly
modified by the presence of another vesicle due to long-range hydrodynamic interaction that is also strengthened by the direct measurements of a single vesicle back-reaction on the velocity field. A direct link to the
rheology of vesicle suspensions still remains a challenge due to the
diversity of the physical parameters and the many-vesicle hydrodynamically-assisted interactions, which could be much stronger than the pair vesicle interaction considered here.
Thus, the experiments provide a lower bound for the interaction scale at $d_{min}/R\approx 3.2\div 3.7$, and so the corresponding upper bound for the volume fraction of non-interacting vesicle suspension is $\phi=0.08\div 0.13$.\\
\indent We are grateful to A. Lishansky and A. Nir for their consulting on colloid particle interactions. This work is partially supported by grants from Israel Science, German-Israeli, and the Minerva Foundations.

\end{document}